\documentclass[aoas]{imsart}

\RequirePackage{amsthm,amsmath,amsfonts,amssymb}
\RequirePackage[authoryear]{natbib}
\RequirePackage[colorlinks,citecolor=blue,urlcolor=blue]{hyperref}
\RequirePackage{graphicx}
\usepackage{xr}
\usepackage{url}
\usepackage[dvipsnames]{xcolor}
\usepackage{array,lscape}
\usepackage{soul}
\usepackage{subcaption}
\usepackage{chngcntr}
\startlocaldefs
\numberwithin{equation}{section}
\theoremstyle{plain}
\newtheorem{theorem}{Theorem}
\theoremstyle{remark}

\makeatletter
\newcommand*{\indep}{%
	\mathbin{%
		\mathpalette{\@indep}{}%
	}%
}
\newcommand*{\nindep}{%
	\mathbin{
		\mathpalette{\@indep}{\not}
	}%
}
\newcommand*{\@indep}[2]{%
	\sbox0{$#1\perp\m@th$}
	\sbox2{$#1=$}
	\sbox4{$#1\vcenter{}$}
	\rlap{\copy0}
	\dimen@=\dimexpr\ht2-\ht4-.2pt\relax
	\kern\dimen@
	{#2}%
	\kern\dimen@
	\copy0 
}
\makeatother

\makeatletter
\newcommand*{\addFileDependency}[1]{
  \typeout{(#1)}
  \@addtofilelist{#1}
  \IfFileExists{#1}{}{\typeout{No file #1.}}
}
\makeatother

\newcommand*{\myexternaldocument}[1]{%
    \externaldocument{#1}%
    \addFileDependency{#1.tex}%
    \addFileDependency{#1.aux}%
}

\myexternaldocument{Supplementary}

\newcolumntype{H}{>{\setbox0=\hbox\bgroup}c<{\egroup}@{}}

\endlocaldefs

\begin{document}

\begin{frontmatter}
\title{Structure learning for zero-inflated counts, with an application to single-cell RNA sequencing data}
\runtitle{Structure learning for zero-inflated counts}

\begin{aug}
\author[A]{\fnms{THI KIM HUE} \snm{NGUYEN}\ead[label=e1]{nguyen@stat.unipd.it}},
\author[B]{\fnms{KOEN} \snm{VAN DEN BERGE}\ead[label=e2]{koenvdberge@berkeley.edu}},
\author[C]{\fnms{MONICA} \snm{CHIOGNA}\ead[label=e3]{monica.chiogna2@unibo.it}},
\and
\author[A]{\fnms{DAVIDE} \snm{RISSO}\ead[label=e4]{davide.risso@unipd.it}}

\address[A]{Department of Statistical Sciences, Universiy of Padova, Padova, Italy, \printead{e1,e4}}
\address[B]{Department of Applied Mathematics, Computer Science and Statistics, Ghent University, Ghent, Belgium and \\ Department of Statistics, University of California, Berkeley, Berkeley, CA, USA, \printead{e2}}
\address[C]{Department of Statistical Sciences, University of Bologna, Italy, \printead{e3}}
\end{aug}

\begin{abstract}
{{The problem of estimating the structure of a graph from observed data is of growing interest in the context of high-throughput genomic data, and single-cell RNA sequencing in particular. These, however, are challenging applications, since the data consist of high-dimensional counts with high variance and over-abundance of zeros. Here, we present a general framework for learning the structure of a graph from single-cell RNA-seq data, based on the zero-inflated negative binomial distribution. We demonstrate with simulations that our approach is able to retrieve the structure of a graph in a variety of settings and we show the utility of the approach on real data.}}
\end{abstract}


\end{frontmatter}

\section{Introduction}\label{intro}
In recent years, a growing interest has developed around the problem of retrieving, starting from observed data, the structure of graphs representing relationships among variables of interest. In fact, reconstruction of a graphical model, known as structure learning, traces back to the beginning of the nineties, and a vast literature exists that considers the problem from various perspectives, within both frequentist and Bayesian approaches (see \citet{doi:10.1146/annurev-statistics-060116-053803} for an extensive review).  But a central role in the renewal of interest on structure learning has been played by molecular biology applications.
In this field, the abundance of data with increasingly large sample sizes, driven by novel high-throughput technologies, has opened the door for the development and application of structure learning methods, in particular applied to the estimation of gene regulatory or gene association networks.

At the inception of transcriptomics, the technology of choice for measuring gene expression was the microarray assay, that, by optically scanning fluorophore intensities, provided data on a continuous scale \citep{irizarry2003exploration}. When it came to (sparse) structure learning from these data, the first proposals assumed that data arose from a multivariate Gaussian distribution, and took advantage of the many results and tools available for such family of distributions (see \citet{schafer2005empirical,WANG2005443,pena2008,yin2011sparse}, among others).

Later, a new technology allowed for the high-throughput sequencing of RNA molecules (i.e., RNA-seq), and quickly established itself as the reference technology for the study of genome-wide transcription levels \citep{wang2009rna}. One of the main advantages of RNA-seq over microarrays is that it allows to analyze small amounts of RNA, making it feasible to study gene expression even at single-cell resolution \citep{kolodziejczyk2015technology}.
This new technology provided statisticians with a wealth of novel problems. Indeed, RNA-seq yields counts, rather than intensities on a continuous scale, as measures of gene expression. Data are usually high dimensional and, typically, come from skewed  distributions with high variance. Moreover, they very often show a large number of zeros, typically larger than expected under a Poisson or negative binomial model \citep{van2013bayesian}.

Structure learning of graphs with such data was initially performed by exploiting data transformations, such as log, Box-Cox, copulas, etc \citep{abegaz2015copula}.
Although data transformation can work well in some circumstances, it can be also ill-suited, possibly leading to wrong inferences in some circumstances \citep{gallopin2013hierarchical}.
Awareness of these problems fueled the development of  methods  for learning (sparse) graphical models tailored to count data.   \citet{allen2013local}, \citet{yang2013poisson} and, more recently, \citet{JMLR:v22:18-401} considered structure learning for Poisson and truncated Poisson counts. A general class of models was studied in \citet{JMLR:v16:yang15a}, which considered graphical models for the class of exponential family.

The challenges posed by RNA-seq technology are exacerbated in single-cell RNA sequencing (scRNA-seq). scRNA-seq allows the measurement of RNA from individual cells, promising to permit the study of gene interactions at an unprecedented resolution \citep{mcdavid2019graphical}.
\textcolor{black}{Some scRNA-seq platforms employ \textit{unique molecular identifiers} (UMIs), which help reduce amplification biases \citep{islam2014quantitative} by counting unique RNA molecules rather than \textit{reads} potentially representing the same molecule more than once. This implies that the distribution of the resulting data is substantially different: read-count data typically show larger counts than UMI data and a more pronounced bi-modality \citep{svensson2020droplet}.}
Moreover, the small amount of RNA present in the cell and the technical limitations of the sequencing platforms (e.g., a limited number of sequenced reads per cell) lead to higher variance and larger fraction of zero counts compared to ``bulk'' RNA-seq \citep{risso2018general,mcdavid2019graphical}. As a result, single-cell RNA-seq gene-wise data distributions are highly skewed and show an abundance of zero counts. Inference using Gaussian models is definitely infeasible even after variance stabilizing transformations and {even} models for count data
{ may suffer from high false discovery rates (see \cite{gallopin2013hierarchical}, and Section \ref{sec:sim}).}
To account for zero-inflation,
\cite{mcdavid2019graphical} proposed a Hurdle model, equivalent to a finite mixture of singular Gaussian distributions. The authors' model, however, does not account for the count nature of the data. 

From this quick tour on problems and methods, it appears evident that principled solutions to structure learning that account for the possibility of over-dispersion and/or zero-inflation are still lacking. 
In this paper, we try to fill this gap. We present a general framework, based on the zero-inflated negative binomial distribution, for learning the structure of a graph from single-cell RNA-seq data. We focus in particular on \textcolor{black}{UMI  data, as its growing popularity suggests that the majority of future studies will employ this technology.}

The remainder of this article is organized as follows. In Section \ref{sec:singlecell} we introduce a motivating dataset; we describe our proposed model in Section \ref{sec: model} and our structure learning procedure in Section \ref{sec:inference}. One key question in the literature is whether zero inflation needs to be accounted for in the data, we offer our perspective in Section \ref{sec:ozi}. After exploring the behavior of our method in simulated data in Section \ref{sec:sim}, we apply our proposal to real single-cell RNA-seq data in Section \ref{sec:realdata}. Section \ref{sec:discussion} concludes the article with a discussion.

\section{A motivating example: single-cell gene expression in the olfactory epithelium}
\label{sec:singlecell}


Despite the distributional challenges described in the previous section, single-cell data offer an unprecedented opportunity to discover cellular dynamics, especially in developing cell populations.
{\color{black} Graphical models could be an important tool to learn gene interactions from single-cell data, to learn how these change across conditions and throughout development, and to identify potentially novel transcription factor target genes.
While graphs are widely used in scRNA-seq to group similar cells in the space of gene expression, our approach learns a graphical model considering genes as nodes. This allows us to model cells as i.i.d. observations from a multivariate distribution in which the genes are the variables and the cells are considered a random sample from the cell population.}

In particular, here, we study gene expression from the mouse olfactory epithelium (OE). This tissue is made of two major mature cell types, olfactory neurons and sustentacular support cells. Furthermore, a stem cell niche provides a mechanism through which the tissue is regenerated \citep{fletcher2017deconstructing,gadye2017injury}. As the aim of the study is to understand how stem cells mature into neurons following tissue damage, we focused only on the cells in the neuronal lineage. Briefly, the olfactory reserve stem cells, called Horizontal Basal Cells (HBC), become activated and subsequently develop into Globose Basal Cells (GBC) and then into immature (iOSN) and finally mature olfactory neurons (mOSN). By reconstructing the structure of the graph for each of these cell types separately, we hope to get a glimpse of the  relationships between genes in neuronal development. {\color{black} For instance, the cell type that results in the most highly connected graph could indicate the most transcriptionally active developmental stage \citep{fletcher2017deconstructing}.}

\section{Model Specification}
\label{sec: model}

\subsection{Preliminaries}

A probabilistic graphical model requires the definition of a pair, $(G,\mathcal{F})$ say. Here, $G =(V,E)$ represents an undirected graph, where $V$ is the set of nodes, and $E = \{(s,t): \,\, s,t \in V, \,\,s \ne t\}$ represents the set of undirected edges. Each node in the graph corresponds to a random variable  $X_s, s \in V;$ the existence of an edge $(s, t) \in E$ indicates the dependency of the random variables $X_s$ and $X_t$. Moreover, $\mathcal{F}$ represents a family of probability measures for the random vector $\mathbf{X}_V,$ indexed by  $V$ and with support $\mathcal{X}_V.$

{
Thanks to the well known Markov properties (global, local, pairwise, see \citet{lauritzengraphical}), the pattern of edges in the graph translates into conditional independence properties for variables in $\mathbf{X}_V,$  which, in turn, allow possible factorizations of $\mathcal{F}$ into smaller, more tractable entities.   In undirected graphical models, each absent edge $(s,t)$ in $E$ has the role of portraying the conditional independence,
$$
X_s\indep X_t|\mathbf{X}_{V\backslash \{s,t\}},
$$
and  the family $\mathcal{F}$ is said to satisfy the pairwise Markov property with respect to $G.$ The smallest undirected graph $G$ with respect to which $\mathcal{F}$ is pairwise Markov is given the name conditional independence graph.

When all variables in $\mathbf{X}_V$ are discrete with positive joint probabilities, as is the case of this paper, the three kinds of Markov properties are equivalent, so that a factorization of the joint probability distribution with respect to the cliques (fully connected subsets of vertices) of the graph $G$ is also guaranteed \citep[][Chap. 3]{lauritzengraphical}.
 }

\subsection{The model specification}
{\color{black}
Let $x_{is}$ be the gene expression for gene $s \in V$ in cell $i\in\{1,2,\ldots,n\}$, we assume that the  distribution of each variable $X_{is},$  conditional to all possible subsets  of variables $\mathbf{X}_{iK}, \, K\subseteq V$ is a zero-inflated negative binomial (zinb) distribution:
\begin{equation}\label{localcondzinb}
    f_{zinb}(x_{is};\mu_{is|K},\theta_s,\pi_{is|K}|\mathbf{x}_{iK\setminus\{s\}})= \pi_{is|K}\delta_0(x_{is})+(1-\pi_{is|K})f_{nb}(x_{is},\mu_{is|K},\theta_{s}|\mathbf{x}_{iK\setminus\{s\}}),
\end{equation}
where $\delta_0(.)$ is the Dirac function, {  $\pi_{is|K}\in [0,1]$ is the probability that a 0 is sampled from a distribution degenerate at zero} and $f_{nb}(.,\mu,\theta)$ denotes the probability mass function of the negative binomial (NB) distribution with mean $\mu$ and inverse dispersion parameter $\theta.$ We assume that
\begin{eqnarray}
	\ln(\mu_{is|K})&=& \nu^{\mu}_{s|K}+\sum_{t\in K\setminus\{s\}} \beta^{\mu}_{st|K}x_{it}\label{link-mu},\\
\text{logit}(1-\pi_{is|K})&=&\nu^{\pi}_{s|K}+\sum_{t\in K\setminus\{s\}} \beta^\pi_{st|K}x_{it}.\label{link-pi}
\end{eqnarray}

A missing edge between node $s$ and node $t$ corresponds to the condition $\beta^{\mu}_{st|K}=\beta^{\mu}_{ts|K}=\beta^{\pi}_{st|K}=\beta^{\pi}_{ts|K}= 0, \,\, \forall K \subseteq V\setminus \{s\}.$ On the other hand, one edge between node $s$ and node $t$ implies that at least one of the four parameters $\beta^{\mu}_{st|K}, \beta^{\mu}_{ts|K}, \beta^{\pi}_{st|K}, \beta^{\pi}_{ts|K}$ is different from 0.
This specification defines a family of models that includes the most common models employed  for count data and embraces a variety of situations.
It is evident that, when 
$\pi_{s|K} = 0, \,\, \forall \ K \subseteq V\setminus \{s\},$ the model reduces to a NB distribution, which, in turn reduces to a Poisson distribution when the inverse dispersion parameter  $\theta_s$ tends to infinity. When $\pi_{s|K} > 0,$ zero-inflation comes into play and zero-inflated Poisson and NB models can be considered. In this case, when $\beta^{\pi}_{st|K} = 0,\,\, \forall \,\,t\in K\setminus\{s\},$  the neighborhood of a node $s$ is defined to be the set of effective predictors of $\mu_{s|K}$ and consists of all nodes $t$ for which $\beta^{\mu}_{st|K} \ne 0.$ On the other side, when $\beta^{\mu}_{st|K} = 0,\,\, \forall\,\, t\in K\setminus\{s\},$  the neighborhood of a node $s$ is defined to be the set of effective predictors of $\pi_{s|K}$ and consists of all nodes $t$ for which $\beta^{\pi}_{st|K} \ne 0.$ In other words, the family includes models in which the structure of the graph is attributable only to one of the two parameter components, $\pi_{s|K}$ or $\mu_{s|K}.$
}

{
The difficulty with our model specification is that the definition of a set of conditional distributions does not guarantee the existence of a valid joint distribution, i.e.,  a  joint distribution that possesses the specified conditionals. This might create difficulties in interpreting the resulting graph in probabilistic terms: if the joint distribution does not exist,  graphical separations stored in $G$ as a result of our model specification might not correspond to conditional independence properties on $\mathcal{F}$. However, our formulation guarantees the existence of the joint distribution in a number of relevant subcases. In the following theorem, we clarify conditions for existence of a joint distribution coherent with the conditional specification { (see Section 1.1,  Supplementary Material \citep{supp}, for a proof).}
}

{
\begin{theorem}\label{theo:jointexist}
Let $\mathbf{X}_V=(X_1, X_2,\ldots,X_p)$ be a $p$-random vector with support $\mathcal{X}_V.$  Assume that a set of univariate conditional probability mass functions of the kind (\ref{localcondzinb}) are given for variables in $\mathbf{X}_V$.  Then, a joint distribution having those conditionals exists if  and only if $\theta_s$ is constant for all  $s  \in V$,  and all regression coefficients $\beta^{\mu}_{st|K}$ 
are negative, $\forall K\subseteq V.$
\end{theorem}

The condition on negativity of local regression coefficients in (\ref{link-mu})  resembles a condition known in the literature of Markov random fields  known as ``competitive relationship'' \citep{besag1974spatial}. Generally speaking, the presence of only negative relations among entities is quite a rare event and incapability of capturing positive dependencies might be a severe drawback in various applications. Nevertheless, the existence of a joint distribution in these specific cases assures that statistical guarantees hold for  conditional approaches to structure learning such as the one used in this paper  and somehow softens the hazard of the use of such algorithms outside the conditions of existence of a joint distribution.
}
\section{Structure Learning}
\label{sec:inference}


A conditional independence graph $G=(V,E)$ on $\mathbf{X}_V$ can be estimated by estimating, for each node $s \in V,$ its neighborhood. Hence, one can proceed by estimating the conditional distribution of  $X_s | \mathbf{X}_{V\setminus s}$ and fixing the neighborhood of $s$ to be the index set of variables $\mathbf{X}_{N(s)}$ on which the conditional distribution depends.

To estimate the neighborhood of each node, we employ the PC-stable algorithm, a variant of the PC algorithm first proposed by \citet{spirtes2000causation}. The PC algorithm starts with a  complete graph on $V.$ Marginal independencies for all pairs of nodes are tested, and edges removed when marginal independencies are found. Then,  for every pair of linked nodes, independence is tested conditional to all subsets of cardinality one of the adjacency sets of the two nodes. This testing procedure is iterated, increasing in turn the size of the conditioning sets, until this reaches its maximum limit, or a limit  imposed by  the user. Reasons for choosing the PC algorithm are many, spanning from its consistency (assuming no latent confounders) under i.i.d. sampling \citep{spirtes2000causation}, to its ability to deal with a large number of variables and only moderately large sample sizes.  The variant that we employ, PC-stable \citep{colombo2014order}, allows to control instabilities due to the order in which  the conditional  independence  tests are performed. {
To perform the tests, deviance test statistics are employed, for which a chi-squared asymptotic distribution can be obtained by standard asymptotic theory.

{\color{black}In what follows, let  $\mathbb{X}=\{\mathbf{x}^{(1)},\ldots,\mathbf{x}^{(n)}\}$ be the collection of $n$ samples drawn from the random vectors $\mathbf{X}_V$, with $\mathbf{x}^{(i)}=(x_{i1},\ldots,x_{ip}),~ i=1,\ldots,n$.
	Starting from the complete graph, for each $s$ and $t\in V\backslash\{s\}$ and  for any set of variables $\mathbf{S}\subseteq \{1,\ldots,p\}\backslash \{s,t\}$, we test,  at some pre-specified significance level, the null hypothesis $H_0: \beta^{\mu}_{st|K}=\beta^{\mu}_{ts|K}=\beta^{\pi}_{st|K}=\beta^{\pi}_{ts|K}= 0$, with $K=\mathbf{S}\cup\{s,t\}$. In other words, we test if the data support the existence of the conditional independence relation $X_s\indep X_t|\mathbf{X}_{\mathbf{S}}$. 
	If the null hypothesis is rejected, there exists an edge $(s,t)$ in the resulting graph. }
	A control is operated on the cardinality of the set $\mathbf{S}$ of conditioning variables, which is progressively increased from 0 to $p-2$ or to $m, \,\, m<(p-2)$.

	Assume $X_s|\mathbf{x}_{K\setminus\{s\}}\sim$ zinb$(X_s;\mu_{s|K},\theta_s,\pi_{s|K}|\mathbf{x}_{K\setminus\{s\}})$, as in Equation \eqref{localcondzinb}.  	The conditional log-likelihood for variable $X_s$ given $\mathbf{x}_{K\setminus\{s\}}$ is obtained by

\begin{eqnarray}\label{smallloglikelihood}
	\ell_s(\boldsymbol{\nu}_{s|K},\boldsymbol{\beta}_{s|K},\theta_s)&=&\sum_{i=1}^n\ln f_{zinb}(x_{is};\mu_{is|K},\theta_s,\pi_{is|K}|\mathbf{x}^{(i)}_{K\backslash\{s\}}),
\end{eqnarray}
where $\boldsymbol{\nu}_{s|K},\boldsymbol{\beta}_{s|K}$ are linked to $\boldsymbol{\pi}_{s|K},\boldsymbol{\mu}_{s|K}$ through Equations \eqref{link-mu} -- \eqref{link-pi}.
	The estimates $\hat{\boldsymbol{\nu}}_{s|K},\hat{\boldsymbol{\beta}}_{s|K},\hat{\theta}_{s}$ of the parameters $\boldsymbol{\nu}_{s|K},\boldsymbol{\beta}_{s|K},\theta_s$  {are obtained  by maximizing the   conditional log-likelihood given in Equation \eqref{smallloglikelihood}}, i.e.,
	\begin{equation*}
	(\hat{\boldsymbol{\nu}}_{s|K},\hat{\boldsymbol{\beta}}_{s|K},\hat{\theta}_{s}) = \text{argmax}_{(\boldsymbol{\nu}_{s|K},\boldsymbol{\beta}_{s|K},\theta_s) \in \mathbb{R}^{2|K|+1} }	\,\,\ell_s(\boldsymbol{\nu}_{s|K},\boldsymbol{\beta}_{s|K},\theta_s).
	\end{equation*}
{ See Section 1.2, Supplementary Material \citep{supp}, for details on the estimation procedure.}
A deviance test statistic for the hypothesis $H_0: \beta^{\mu}_{st|K}=\beta^{\pi}_{st|K}= 0$  can be obtained as
	$$D_{s|K}=2(\ell_s(\hat{\boldsymbol{\nu}}_{s|K},\hat{\boldsymbol{\beta}}_{s|K},\hat{\theta}_{s})-\ell_s(\hat{\boldsymbol{\nu}}^0_{s|K},\hat{\boldsymbol{\beta}}^0_{s|K},\hat{\theta}^0_{s})),$$
where $\hat{\boldsymbol{\nu}}^0_{s|K},\hat{\boldsymbol{\beta}}^0_{s|K},\hat{\theta}^0_{s}$ are the maximum likelihood estimates of the parameters under $H_0.$
It is readily available that $D_{s|K}$ is asymptotically chi-squared distributed with 2-degrees of freedom under the null hypothesis, provided that some general regularity conditions hold.

{\bf Remark 1.}
 On assuming faithfulness of the node conditional distributions to the graph $G,$ consistency of the algorithm can be proved in the case of  competitive relationships in $\mu_s|K$  by suitably modifying results in \citet{JMLR:v22:18-401}. We recall that a distribution  $P_\mathbf{X}$ is said to be faithful to the graph $G$ if for all disjoint vertex sets $A,B,C \subset V$ it holds
 $$\mathbf{X}_A\indep \mathbf{X}_B|\mathbf{X}_C\Rightarrow A\indep_G B|C,$$
 where $A\indep_G B|C$ means that $A$ and $B$ are separated in $G$ by $C.$
Thanks to the equivalence between local and global Markov properties, faithfulness of the local distributions guarantees faithfulness of the joint distribution.

{\bf Remark 2.}
Although a theoretical proof of convergence of the algorithm is in question in the case of unrestricted relationships among variables, inference on the structure is still principled within a pseudo-likelihood perspective, i.e.,  by approximating the likelihood function by a product of the conditional likelihood functions. Different pseudo-likelihood-based structure estimators have been shown to be consistent under a conditional model construction (see, \cite{10.1214/009053605000000912}, among others). See also \citet{JMLR:v22:18-401} for an empirical exploration of convergence of a similar algorithm under the Poisson assumption in the case of unrestricted relationships among variables.

{
{\bf Remark 3.}
 A large sample size, as typical in the applications at hand, impacts on the actual significance level of individual tests. Moreover, a multiplicity of tests are performed by the algorithm. For this reason, we advice to set the nominal level of the test $\alpha$ to $\alpha_n=2(1-\Phi(n^b)),$ where $0<b<1/2$ is related to the average neighborhood size. This choice is based on results in \citet{JMLR:v22:18-401} and guarantees that the probability that a type I or II error occurs in the whole testing procedure goes to zero as $n \rightarrow \infty,$ i.e., it asymptotically controls the family-wise error rate of all potential tests that could be done. 

{\bf Remark 4.}
The chosen learning strategy has some advantages over alternative approaches based on sparse regressions (see also \citet{JMLR:v22:18-401} for an extended discussion in the Poisson case). Sparsity can be easily implemented by a control on the conditional set size, instead of a control on parameter magnitudes, which can lead to over-shrinkage.  Moreover, it offers computational advantages, especially when sparse networks are the target of inference.
}}

\section{Zero inflation: a real issue?}\label{sec:ozi}
The need for modeling zero-inflation in single cell data is a question at the core of an ongoing debate, with several authors arguing that the negative binomial distribution is sufficient to fit single-cell RNA-seq data when unique molecular identifiers are used \citep{vieth2017powsimr,townes2019feature,svensson2020droplet,sarkar2021separating}.
Indeed, the ability to distinguish between a non zero-inflated distribution and  zero-inflated alternatives highly depends on the relative size of the parameters of the distributions.

To gain a better understanding of this problem, we have tried to assess the misspecification cost due to assuming a zero-inflated distribution when no zero inflation occurs. To this aim, we confined ourselves to a univariate case with no covariates, fixed a non zero-inflated model and measured the model misspecification cost occurring when using its zero-inflated counterpart by using the squared Hellinger distance as loss function. Such a loss function should, in principle, indicate, in an inferential sense, how far apart the two distributions are.

{
To this aim, let $\mathcal{Y} =\{0,1,2,\ldots,+\infty\}$ be the support of a discrete variable $Y$. We consider for $Y$ a true probability distribution $P(y; \phi_0),\,\,\phi_0\in\Phi$, as well as a family $\mathcal{F} = \{Q(y; \psi),\,\,\psi\in\Psi\},\,\,\Phi \subseteq \Psi,$ of zero-inflated versions of the true probability distribution $P(y; \phi_0).$  The  squared Hellinger distance between two probability distributions $P$ and $Q$ is defined as
\begin{eqnarray*}
d_h^2(P,Q)&=&\frac{1}{2}\sum_{y \in \mathcal{Y}}(\sqrt{p}_y-\sqrt{q}_y)^2\\
&=& 1-\sum_{y \in \mathcal{Y}}\sqrt{p_yq_y},
\end{eqnarray*}
where $p_y=P(y; \phi_0)$ and $q_y=Q(y; \psi).$

In particular, assume that $P$ is a NB distribution with $\phi_0=(\mu_0, \theta_0)$ and $Q$ is a zinb distribution, defined as $$Q(y;\psi)=\pi\delta_0(0)+(1-\pi)P(y; \phi_0),$$
with $\psi=(\phi_0, \pi).$
Hence, the squared Hellinger distance of $P$ and $Q$ can be written as
\begin{eqnarray*}
d_h^2(P,Q)&=& 1-\sqrt{1-\pi}(1-P(0; \phi_0))-
                         \sqrt{P(0; \phi_0)}\sqrt{\pi+(1-\pi)P(0; \phi_0)}.
\end{eqnarray*}

{\color{black}Figure~\ref{Hellingerdist-fig}, and Supplementary Table S1 show the value of the Hellinger distance in a number of cases.} As expected,  the  distance increases with the probability of zero inflation $\pi.$ However, when the inverse dispersion parameter $\theta_0$ and/or the mean $\mu_0$ are small, the distance between the distributions is small even in the case of moderate to large $\pi$. In fact, when $\mu_0$ and $\theta_0$ are both small (low mean and high variance) the two distributions are close even when $\pi = 0.9$.

\begin{figure}[htbp]
\centering
\vspace*{-1cm}
\includegraphics[width = 1\linewidth, height=0.6\textheight]{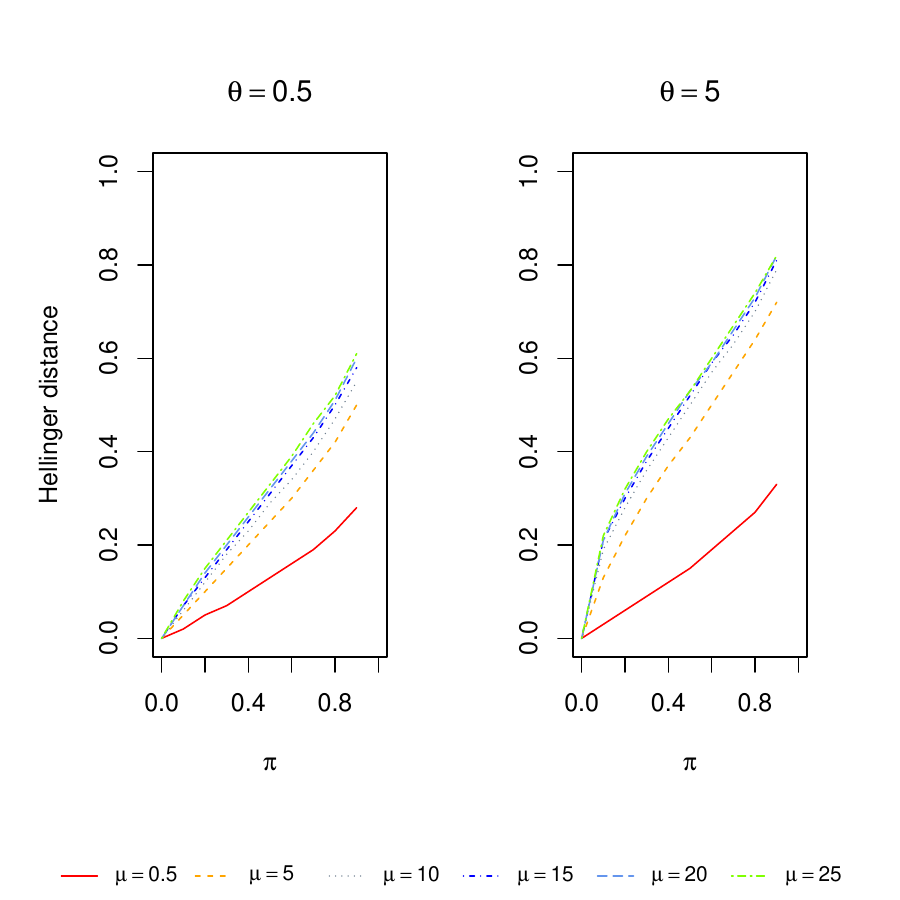}
\caption{Hellinger distance between zinb and NB distribution.}
\label{Hellingerdist-fig}
\end{figure}

As, broadly speaking, maximum likelihood estimators and minimum Hellinger distance estimators are asymptotically equivalent, it emerges that, in inferential terms, the degree of zero inflation of a true model could be difficult to ascertain, as suitable choices of the parameters of the non contaminated component may possibly absorb the excess of zeros generated by the contamination. { This is despite identifiability of the zinb model (see Section 1.3, Supplementary Material \citep{supp}, for a proof)}.  These remarks might contribute to the ongoing debate about existence of zero-inflation from a novel perspective.

}
\section{Simulations}
\label{sec:sim}
We devote this section to the empirical study of consistency of the proposed algorithms.
In particular, we concentrate on the ability of  proposed methods  to recover the true structure of the graphs. We also list the running time of each algorithm.  As measures of the test's accuracy, we  adopt three criteria including Precision $P$;  Recall $R$; and their harmonic mean, known as $F_1$-score, respectively defined as
$$ P=\frac{TP}{TP+FP},\, R=\frac{TP}{TP+FN},\, F_1=2  \frac{P .  R}{P+R},$$
where TP (true positive), FP (false positive), and FN (false negative) refer to the { number of} inferred edges \citep{liu2010stability}.  

The considered algorithms  are listed below, along with specifications, if needed, of tuning parameters. For all PC-like algorithms, we let the maximum cardinality of conditional independence set be $m=8$ for $p=10$ and $m=3$ for $p=100$.

	\begin{itemize}
		\item[-] {\bf PC-zinb1:}  zinb models in which  the structure of the graph is attributable to both of the two parameter components $\mu_{s|K}$ and $\pi_{s|K}$;
		\item[-] {\bf PC-zinb0:} zinb models in which  the structure of the graph is attributable to only the parameter component $\mu_{s|K}$ and  consider $\pi_{s|K}$ as a constant (i.e., $\beta^{\pi}_{st|K} = 0,\,\, \forall\,\, t\in K\setminus\{s\}, \, \forall\,\, s\in V$ );
			\item[-] {\bf PC-nb:}  Negative binomial model, i.e., the special case of zinb models where $\pi_{s|K}=0$;
		\item[-] {\bf PC-pois:}  Poisson model \citep{JMLR:v22:18-401}.

	\end{itemize}

\subsection{Data generation}
For two different cardinalities ($p=10$ and $p=100$),  we consider three graphs of different structure: (i) a scale-free graph, in which the node degree distribution follows a power law; (ii) a hub graph, where each node is connected to one of the hub nodes; (iii) a Erdos-Renyi graph, where the presence of the edges 
is drawn from   {independent and identically distributed} Bernoulli random variables.

		To construct the scale-free and Erdos-Renyi networks, we employed the R package \textit{igraph} \citep{csardi2006igraph}. For the scale-free networks, we followed the Barabasi-Albert model with constant out-degree of the vertices $\nu=2$ for $p=10$ and $\nu=0.2$ for $p=100$. For the Erdos-Renyi networks, we followed the Erdos-Renyi model with probability to draw one edge between two vertices $\gamma=0.3$ for $p=10$ and $\gamma=0.03$ for $p=100$. To construct the hub networks, we assumed 2 hub nodes for $p= 10$, and 5 hub nodes for $p=100$.  See Supplementary Figure S2 and Supplementary Figure S3 for representative plots of the three chosen graphs for $p=10$ and $p=100$, respectively.

For the given graphs, 50 datasets were sampled with four different sample sizes, $n=\{100,200,$ $500,1000\}$ for $p=10$, and three different sample sizes, $n=\{200,500,1000\}$ for $p=100$.  	To generate the data, we followed the approach of the Poisson models  in \citet{allen2013local}.
	Let $\mathbb{X}\in\mathbb{R}^{n\times p}$ be the set of $n$ independent observations of random vector $\mathbf{X}$.
		Then, $\mathbb{X}$ is obtained from the following model
	$\mathbb{X}=\mathbb{Y}A+\mathbb{\epsilon},$
	where $\mathbb{Y}=(y_{st})$ is an $n\times (p+p(p-1)/2)$ matrix whose entries $y_{st}$ are realizations of independent random variables $Y_{st}\sim$ zinb$(\mu, \theta, \pi)$ (or NB$(\mu, \theta)$; or Pois$(\mu))$ and $\mathbb{\epsilon}=(e_{st})$ is an $n\times p$ matrix with  entries $e_{st}$ which are realizations of random variables $E_{st}\sim$ zinb$(\mu_{nois}, \theta, \pi)$ (or nbinom$(\mu_{nois}, \theta)$; or Pois$(\mu_{nois}))$.
	This approach leverages the additive property of these distributions and allows us to generate the required dependencies. In particular, let $B$ be the adjacency matrix of a given true graph, then  $A$ takes the following form  $A=[I_p;P\odot(1_p tri(B)^T)]^T$. Here, $P$ is a $p\times (p(p-1)/2)$ pairwise permutation matrix, $\odot$ denotes the elementwise product, and $tri(B)$ is the $(p(p-1)/2)\times 1$ vectorized upper triangular part of $B$ {\citep{allen2013local}}. 

\subsection{Results}

\begin{figure}[htbp]
\centering
\includegraphics[width = 1\linewidth, height=0.9\textheight]{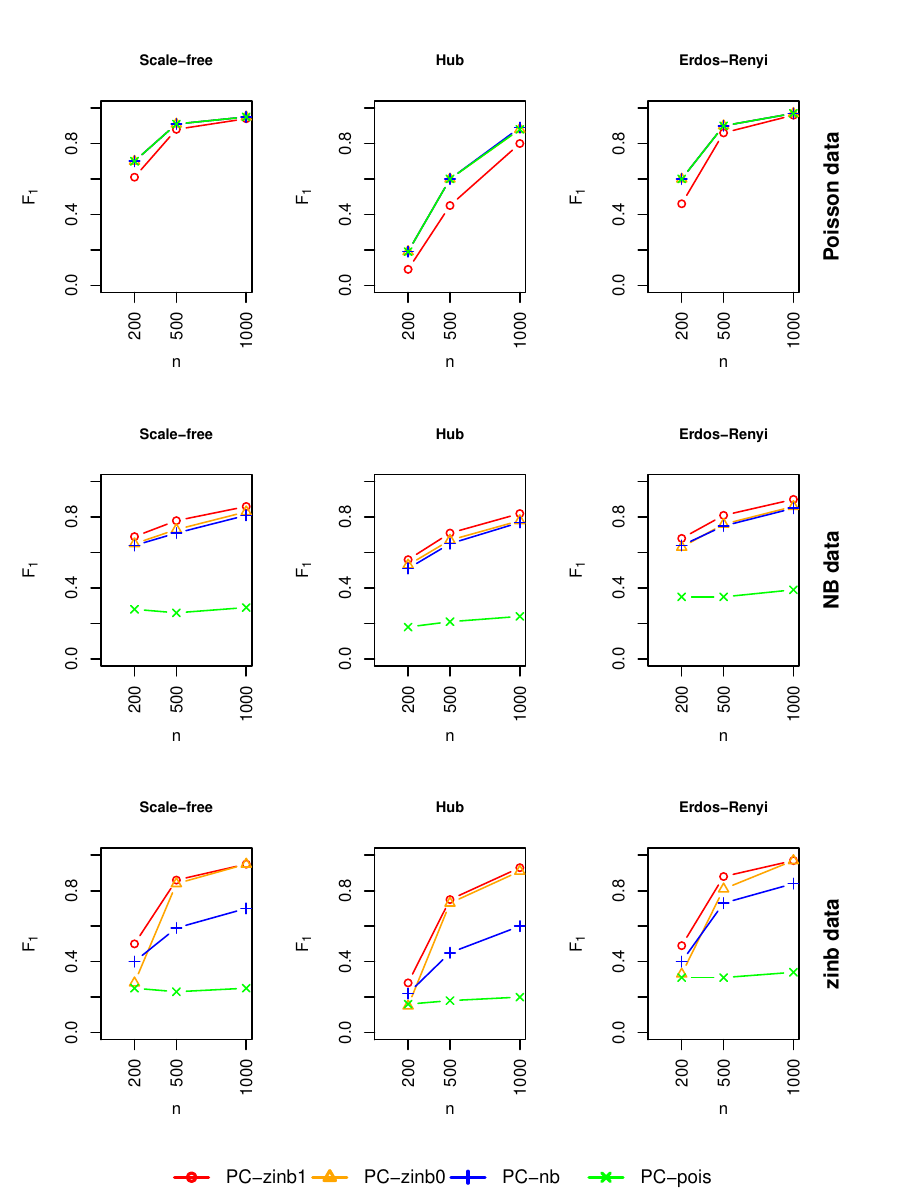}
\caption{$F_1$-score of the considered algorithms for the three types of graphs in Supplementary Figure S3 with $p=100,\mu=5,\theta=0.5, \pi=0.7$: scale-free; hub; Erdos-Renyi. The data were simulated from Poisson (top), NB (middle), and zinb (bottom) models. PC-zinb1:  zinb model in which  the structure of the graph is attributable to both of the two parameter components $\mu_{s|K}$ and $\pi_{s|K}$; PC-zinb0: zinb model in which  the structure of the graph is attributable to only the parameter component $\mu_{s|K}$ and $\pi_{s|K}$ is constant; PC-nb: Negative binomial model, i.e., the special case of zinb models where $\pi_{s|K}=0$; PC-pois:  Poisson model of \citet{JMLR:v22:18-401}.}
\label{graph100-mu5}
\end{figure}
\begin{figure}[htbp]
\centering\includegraphics[width = 1\linewidth, height=0.9\textheight]{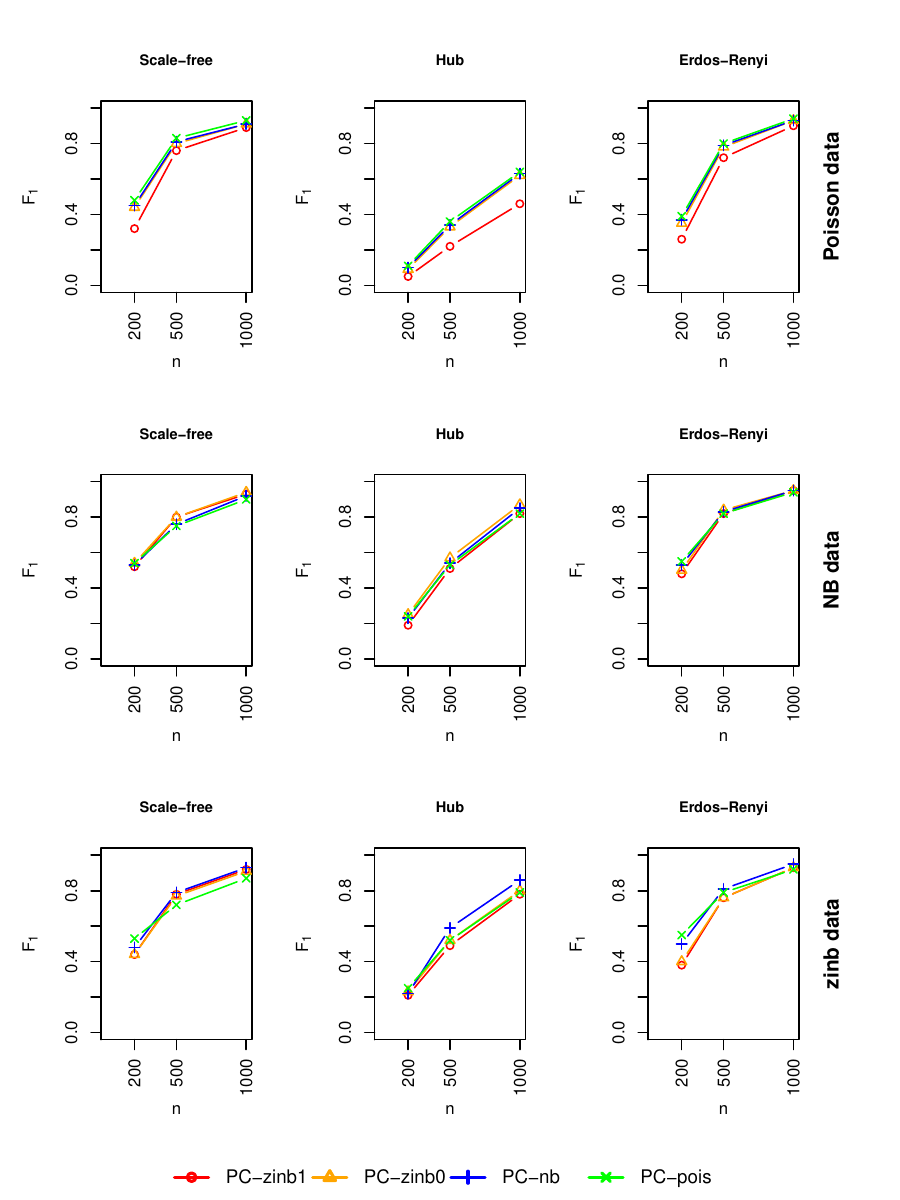}
\caption{$F_1$-score of the considered algorithms for the three types of graphs in Supplementary Figure S3 with $p=100,\mu=0.5,\theta=0.5,\pi=0.7$: scale-free; hub; Erdos-Renyi. The data were simulated from Poisson (top), NB (middle), and zinb (bottom) models. PC-zinb1:  zinb model in which  the structure of the graph is attributable to both of the two parameter components $\mu_{s|K}$ and $\pi_{s|K}$; PC-zinb0: zinb model in which  the structure of the graph is attributable to only the parameter component $\mu_{s|K}$ and $\pi_{s|K}$ is constant; PC-nb: Negative binomial model, i.e., the special case of zinb models where $\pi_{s|K}=0$; PC-pois:  Poisson model of \citet{JMLR:v22:18-401}.}
\label{graph100-mu05}
\end{figure}

{Figures \ref{graph100-mu5} and \ref{graph100-mu05} show the Monte Carlo means of the $F_1$-scores for each of the considered methods with $p=100$ and low signal-to-noise ratio ($\mu_{noise}=0.5$), at high ($\mu=5$) and low ($\mu=0.5$) mean levels, respectively.
Each value is computed as the average of  the 50 values obtained by simulating 50 samples for the model corresponding to each network.
Monte Carlo means of Precision $P$, Recall $R,$ and $F_1$-score are given in Supplementary Tables S2--S4.}

{The two values of $\mu = \{5, 0.5\}$ were chosen to mimic typical values observed in real full-length and droplet-based datasets, respectively. In fact, the mean expression level of transcription factors in the dataset presented in Section \ref{sec:singlecell} is $0.67$ (median $0.14$), while the mean expression level of transcription factors in a similar experiment performed with a full-length protocol \citep{fletcher2017deconstructing} is $32.03$ (median $7.89$).}

These results indicate that the PC-zinb1 algorithm and its variants (PC-zinb0, PC-nb, PC-pois) are consistent in terms of reconstructing the structure from given data.
{In fact, when the model is correctly specified, the $F_1$-scores of the algorithms are close to 1 when $n\ge 1000$ in all scenarios. This means that the proposed algorithm is able to recover the underlying graph from the given data for both low (Fig. \ref{graph100-mu05}) and high (Fig. \ref{graph100-mu5}) mean levels.}

{When the data are generated with a high mean level ($\mu=5$), the PC-pois algorithm performs well only when it is the true model, i.e., for data generated from Poisson random variables (Fig. \ref{graph100-mu5}; Supplementary Table S4; Supplementary Figs.S10 and S12). In the other scenarios, PC-pois often shows a low Precision (Fig. \ref{graph100-mu5}; Supplementary Tables S2 and S3; Supplementary Fig. S10). This result is expected since the node conditional Poisson distributions are unable to model the over-dispersion generated by the (zero-inflated) negative binomial distributions.}

{On the other end of the spectrum, the more general zinb models work well in all scenarios (Fig. \ref{graph100-mu5}; Supplementary Tables S2 -- S4; Supplementary Figs.S10 and S12). This is not surprising as the data are generated according to models (e.g, Poisson, NB) that can be seen as special cases of the zinb distribution, which means that in all tested scenarios the zinb model is correctly specified.}

{The PC-nb algorithm, based on the negative binomial assumption, performs reasonably well (Fig. \ref{graph100-mu5}; Supplementary Tables S2 -- S4; Supplementary Figs. S10 and S12). However, in the hub graph (center column of Fig. \ref{graph100-mu5}), its performances are slightly worse than the zinb models, showing low Precision when the true data generating distribution is node conditional zinb (Supplementary Fig. S10; Supplementary Table S2). This result indicates that a zero inflated negative binomial model may be needed when the mean is large \citep{risso2018general}.}

{As we expected from the considerations reported in Section \ref{sec:ozi}, the performances of the variants of PC-zinb are quite similar to each other when the mean and the dispersion parameter are both small, i.e., when the data are characterized by low mean and high variance ($\mu=0.5, \theta=0.5$; Fig. \ref{graph100-mu05}; Supplementary Tables S2 -- S4; Supplementary Figs. S11 and S13). This might be explained by the fact that a suitable choice of the parameters may allow non-zero inflated models to absorb the excess of zeros (see Section \ref{sec:ozi} for more details). Therefore, when applying our approach on this type of data, one should use the simplest variant, (i.e., PC-pois) to leverage the better computational performance (see last column of Supplementary Table S2 -- S4).
}

{Moreover, we see no difference in the performance of the PC-zinb variants (PC-zinb1 and PC-zinb0). This is perhaps not surprising, as we simulated the same structure of the graph for both $\mu$ and $\pi$. These results suggest that the information inferred from $\mu$ is sufficient to reconstruct the correct graph in this case.}


{Finally, we compare the results to those obtained with the algorithm of \citet{mcdavid2019graphical}, which employs a Gaussian Hurdle model (see Supplementary Table~S2--S4). The Hurdle model, applied to log transformed data shifted by 1, performs reasonably well only with a sufficient sample size ($n\ge 1000$) in the case of Erdos-Renyi and scale-free graphs, but is unable to correctly reconstruct the hub graphs even at large sample sizes. An extensive analysis of the results of the hub graph case revealed that the graph recovered by the Hurdle model is almost empty in a number of cases, especially at low sample sizes.}

{We have focused here on $p=100$, as this setting is closer to our real application. The results for $p=10$ are reported in Supplementary Figures S4--S9 and Supplementary Tables S5--S7 and lead to similar conclusions.}

{\color{black}
\section{Results on real data}
\label{sec:realdata}

We demonstrate our method on the motivating example dataset described in Section \ref{sec:singlecell}.
To this aim, we analyzed a set of cells, assayed with 10X Genomics (v2 chemistry) after injury of the  OE, to characterize HBCs  and their descendants during regeneration \citep{brann2020non}. Starting from an initial set of 25{,}469 cells, low-quality samples as well as potential doublets were removed as described in \cite{brann2020non}. After clustering with the Leiden algorithm \citep{Traag2019}, known marker genes were used to identify cell types. We discarded the cell types outside of the neuronal lineage (macrophages, sustentacular cells, and microvillar cells), obtaining a dataset consisting of 7782 HBCs, 5418 activated HBCs (HBC*), 755 GBCs, 2859 iOSN, and 929 mOSN. For more details on the data preprocessing \textcolor{blue}{and cell annotation}, see \cite{brann2020non}.

We perform two complementary analyses on two different subsets of the dataset. First, we focus on transcription factor (TF) genes, with the aim of identifying important networks of regulation in the different cell types that constitute the neuronal developmental lineage. We then turn our attention to the activated HBCs, a critical stage of neurogenesis, with the aim of identifying important transcription factors that regulate genes important for stem cell differentiation.

}
\subsection{Transcription factor genes}

Our first analysis focuses on the total set of 1543 known transcription factors in mouse, which are thought to regulate the observed differentiation processes. We furthermore focus on the differentiation path starting at the HBC* stage (i.e., activated stem cells upon injury) up to mature neurons, therefore investigating the entire neuronal lineage in the trajectory of this dataset. As previously discussed in Section \ref{sec:singlecell}, we expect four different cell types along this path, being respectively HBC*, GBC, iOSN and mOSN, and we estimate the structure of the graph for each of these cell types. \textcolor{black}{We selected the top 1000 cells with the highest means from the cell types that had more than 1000 cells (HBCs, HBC*, iOSN) to ensure a fair comparison between groups. In fact, the power of our algorithm to detect edges increases with the sample size and since one of the goals of this analysis is to compare the graphs across cell types we want to avoid a confounding effect due to the number of cells. See Supplementary material \citep{supp}, Section 2, for details on the preprocessing.}

The average degree of the graphs is highest at the activated stem cell stage, with an average degree of $4$, and decreases as cells develop to mature neurons, with average degrees of $3.9, 3.3$ and $3.5$ for the GBC, iOSN and mOSN networks, respectively. To interpret the graph structure, we focus on the 2-core of each network, i.e., we retain TFs that are associated with at least two other TFs, a preprocessing step that helps in understanding the core structure \citep{wang2016}.

We identify communities in each graph using the Leiden algorithm \citep{Traag2019} and, in order to validate the associations discovered by PC-zinb, we interpret each of the communities by computing overlaps with known functional gene sets in the MSigDB database \citep{Subramanian2005, Liberzon2015}, see Supplementary Material \citep{supp} Section 2 for details. The interpretation of these communities relies on known processes involved in the development of the olfactory epithelium as found by previous research (e.g., \citet{fletcher2017deconstructing, gadye2017injury}).

\begin{figure}[htbp]
\centering
  \begin{subfigure}[b]{.5\linewidth}
    \centering
    \includegraphics[width=1.3\textwidth, height=0.4\textheight]{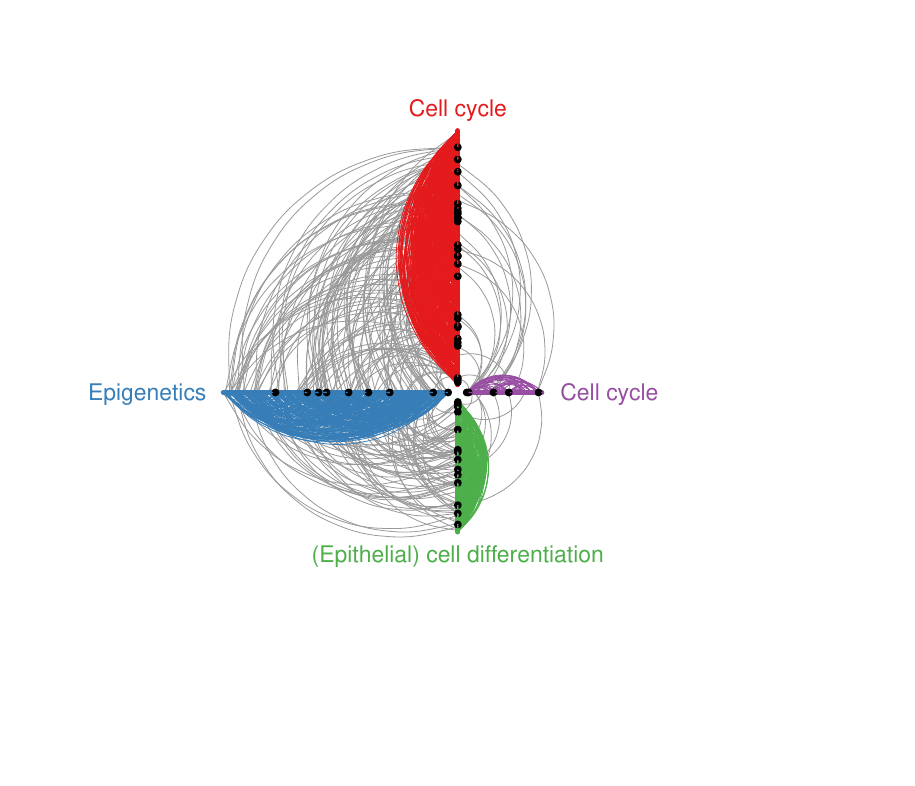}
    \vspace{-2cm}
    \caption{HBC*}\label{fig:1a}
      \vspace{1cm}
  \end{subfigure}%
  \begin{subfigure}[b]{.5\linewidth}
    \centering
    \includegraphics[width=1.3\textwidth, height=0.4\textheight]{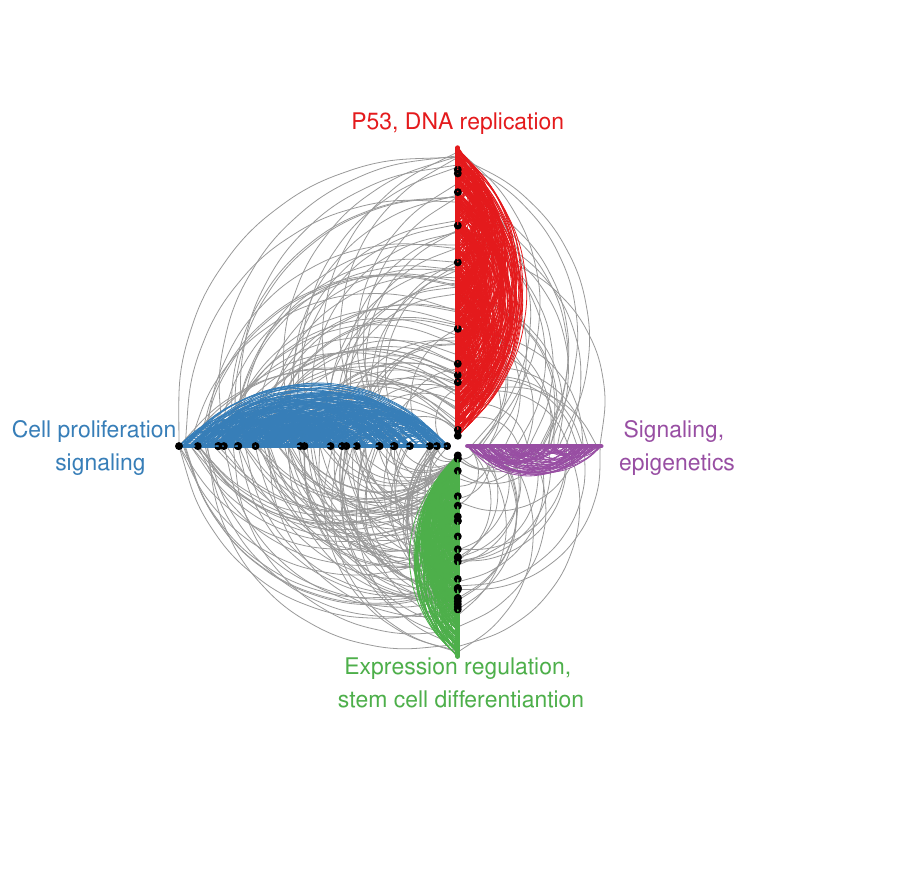}
    \vspace{-2cm}
    \caption{GBC}\label{fig:1b}
      \vspace{1cm}
  \end{subfigure}\\
  \begin{subfigure}[b]{.53\linewidth}
    \centering
    \includegraphics[width=1.13\textwidth, height=0.4\textheight]{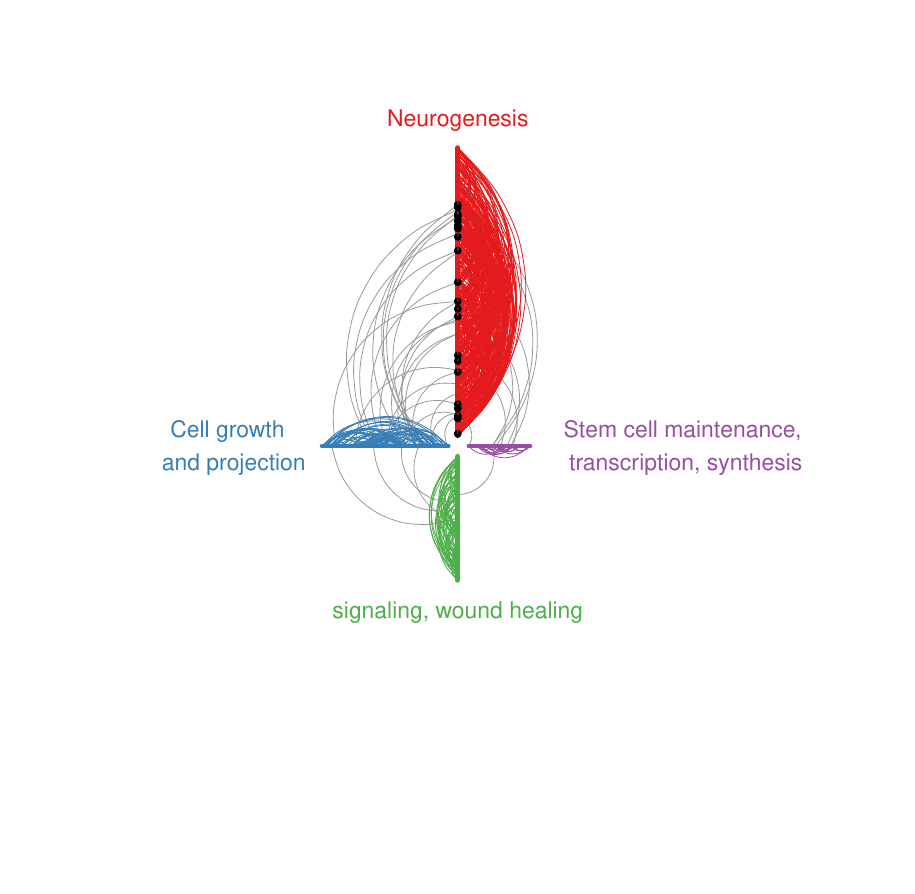}
    \vspace{-2cm}
    \caption{iOSN}\label{fig:1c}
  \end{subfigure}%
  \begin{subfigure}[b]{.5\linewidth}
    \centering
    \includegraphics[width=1.3\textwidth, height=0.4\textheight]{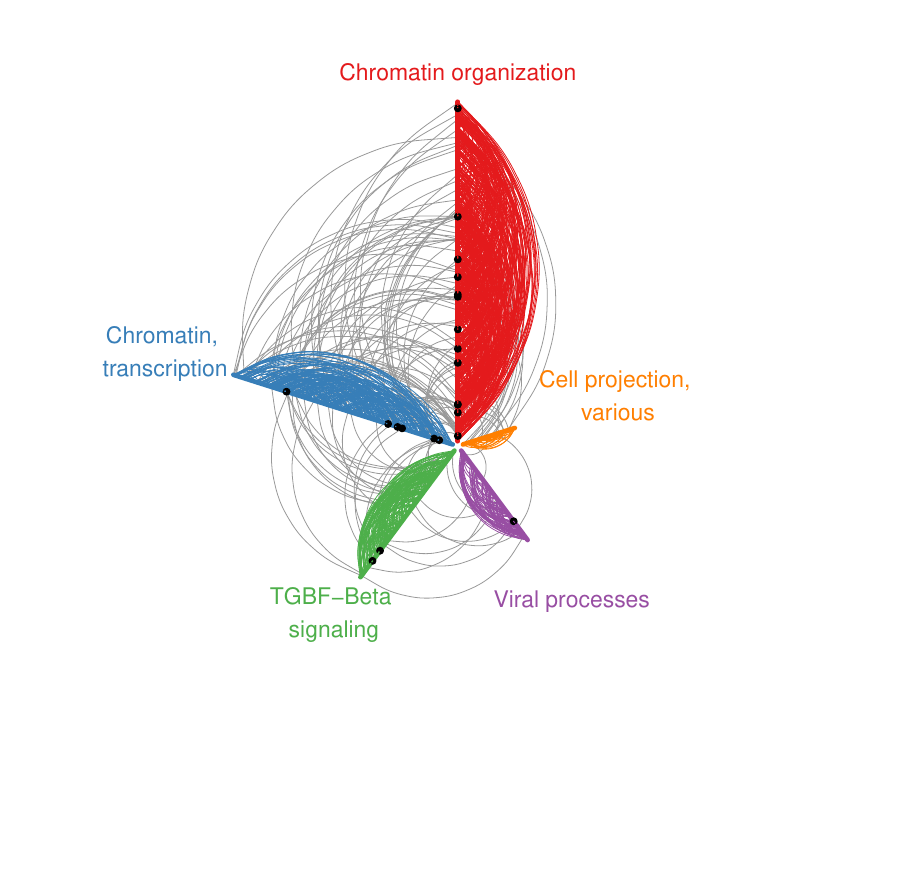}
    \vspace{-2cm}
    \caption{mOSN}\label{fig:1m}
  \end{subfigure}%
  \caption{Hive plots \citep{krzywinski2012hive,hiver} of TF gene networks estimated with PC-zinb. Gene communities were estimated using the Leiden algorithm and are represented on the axes of the plots and by different edge colors. The length of each axis is proportional to the size of the corresponding community; {edges between two nodes in the same community are drawn in a community-specific color, while edges between two nodes in two different communities are colored in gray}; hub nodes, \textcolor{black}{defined as nodes with more than 9 neighbors,} are represented as solid black circles. Each axis (community) was annnotated with the most enriched gene set (see Supplementary Material Section 2 \citep{supp}).}
\label{fig:hiveplot}
\end{figure}

In the HBC* cell type, cells have been injured $\sim 24$h ago, so we expect response to injury, and stem cells actively preparing for differentiation, as well as replication to produce more stem cells to repair the epithelium. Four communities are discovered in the association network (Figure \ref{fig:hiveplot}), broadly involved in either cell cycle, epigenetic mechanisms and (epithelial) cell differentiation (Supplementary Table S9). These communities reflect the need to divide in order to produce more cells, epigenetic mechanisms that are likely required to activate molecular processes upon injury, and the differentiation of stem cells to restore the damaged epithelium.

In the GBC cell type, we expect cells to proliferate to produce immature neurons. We discover four communities (Figure \ref{fig:hiveplot}), broadly involved in DNA replication, cell proliferation, signaling, expression regulation and cell differentiation (Supplementary Table S10). Relevant pathways, such as the P53 and notch signaling pathways, are also recovered for specific communities, and have previously been found to be involved in neurogenesis in neuroepithelial stem cells \citep{MarinNavarro2020, Wang2011}.

In the immature olfactory sensory neuron (iOSN) stage, we expect basal cells to start developing into immature neurons. Four communities are discovered (Figure \ref{fig:hiveplot}), of which one community comprises the majority of the graph, i.e., $63\%$ of all TFs retained in the graph, and importantly is involved in neurogenesis (Supplementary Table S11). Other, smaller, communities are enriched in processes such as cell and axon growth, wound healing, signaling and cell population maintenance.

Finally, in the mature olfactory sensory neuron stage (mOSN), we expect the final differentiation to functional neurons. Five communities are discovered (Figure \ref{fig:hiveplot}), again with very different sizes. The largest communities are enriched in broader processes related to chromatin organization and transcription, possibly reflecting the basic changes required for cells to develop into and maintain at the mature stage (Supplementary Table S12). The third largest community is enriched specifically in the TGF-Beta pathway, known to be required for neurogenesis, and to modulate inflammatory responses \citep{Meyers2017}.

Taken together, these results confirm previously known processes associated with differentiation of HBCs into mature neurons upon injury, with relevant processes highlighted by communities of transcription factors. Furthermore, while the community detection results are useful to validate the estimated graphs, they also provide a gateway to more detailed analysis, e.g., investigation of hub genes (e.g., \citet{Chen2018}) or master regulators of development (e.g., \citet{Sikdar2017}), therefore unlocking powerful interpretation of single-cell RNA-seq datasets. \textcolor{black}{We give an example of such detailed analysis in the next paragraph, in which we focus on the role of the Trp63 TF in activated HBCs.}

{\color{black}
\subsection{Stem cell differentiation}
Our second analysis focuses on a set of 242 genes, annotated with the term ``stem cell differentiation'' in the Mouse Genome Database \citep{bult2019mouse}, expressed in the activated HBC cell type. Following the same preprocessing employed for the first analysis, and detailed in Section 2 of the Supplementary Material \citep{supp}, we obtain a dataset consisting of 1000 cells and 160 genes.

Our goal here is to infer the interactions among genes, with a particular focus on the role of TFs in regulating target genes. Importantly, in this second analysis, we include many genes that are not TFs, allowing us to focus on which genes are regulated by TFs at this specific point in development.

We expect to find several TFs as hub nodes in the graph. In fact, hub nodes, i.e., nodes with a particularly high number of connections, may represent sites of signaling convergence, potentially indicating those genes that regulate other genes.

The PC-zinb algorithm inferred a sparse graph, shown in Figure \ref{fig:HBCactstem}, where hub nodes are displayed with a circle or a hexagon (when they are TFs). It is immediate to recognize important TFs previously demonstrated to be involved with stem cell differentiation, e.g., \textit{Trp63} \citep{senoo2007p63}, \textit{Sox2} \citep{liu2013multiple}, and \textit{Sox9} \citep{jo2014versatile}. Other hub nodes include genes that, while not TFs, have been shown to play a central role in this biological process. For instance, \textit{Epcam} is known to be essential for the maintenance of self-renewal in stem cells \citep{gonzalez2009epcam}. Another example is \textit{Ptn}, the gene encoding the pleiotrophin growth factor, which has significant roles in cell growth and survival and has been demonstrated to be essential for stem cell maturation and neuronal development \citep{tang2019neural}.

\begin{figure}[htbp]
\centering
    \includegraphics[width=1\textwidth]{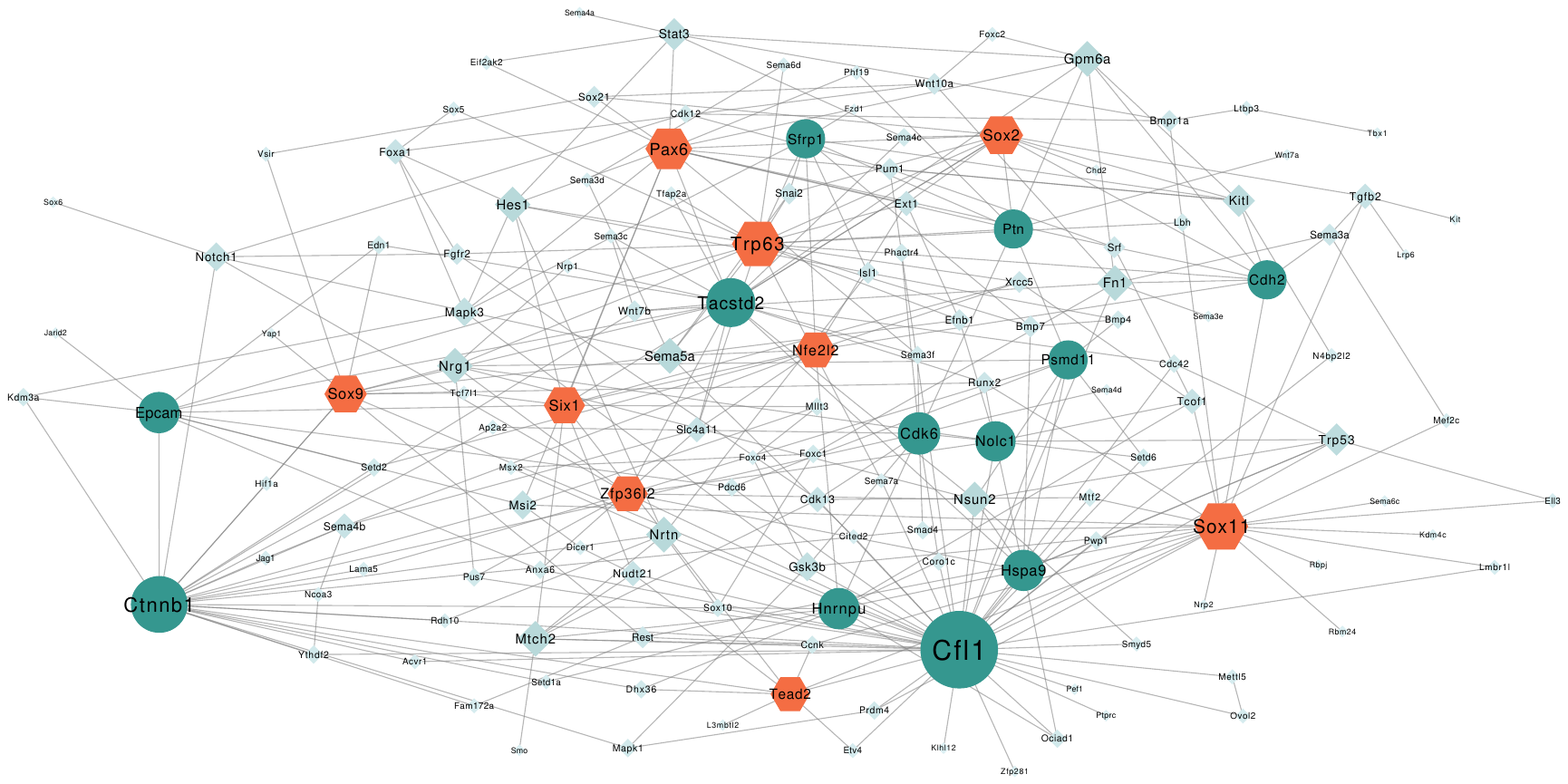}
    \vspace{-2cm}
    \caption{Network of Stem cell differentiation gene set estimated with PC-zinb. Hub nodes are displayed with a circle, hub nodes that are also TF genes are displayed with a hexagon, and the remaining nodes are displayed with a diamond.}\label{fig:HBCactstem}
\end{figure}%

We next focus on one of the most important TFs for stem cell activation, \textit{Trp63}, by zooming in the sub-network made of this gene and its direct neighbors (Fig. \ref{fig:HBCactTrp63trans}). \textit{Trp63} is one of the most important hubs in the network inferred by PC-zinb, with 20 connections. To validate the biological meaning of these connections, we leverage existing external data. In particular, \citet{riege2020dissecting} performed a meta-analysis of 20 publicly available Chromatin Immunoprecipitation (ChIP-seq) datasets to create a curated catalog of p63 (the human ortholog of \textit{Trp63}).
Out of the 20 direct targets of \textit{Trp63} in our network, 15 have been confirmed by \citet{riege2020dissecting} as direct targets of p63, i.e., there is experimental evidence that the p63 protein binds either at the transcription start site (TSS) or upstream, indicating that p63 is either a promoter or enhancer of these genes \citep[Supplementary File 3 of][]{riege2020dissecting}.

\begin{figure}[htbp]
\centering
    \includegraphics[width=1\textwidth, height=0.4\textheight]{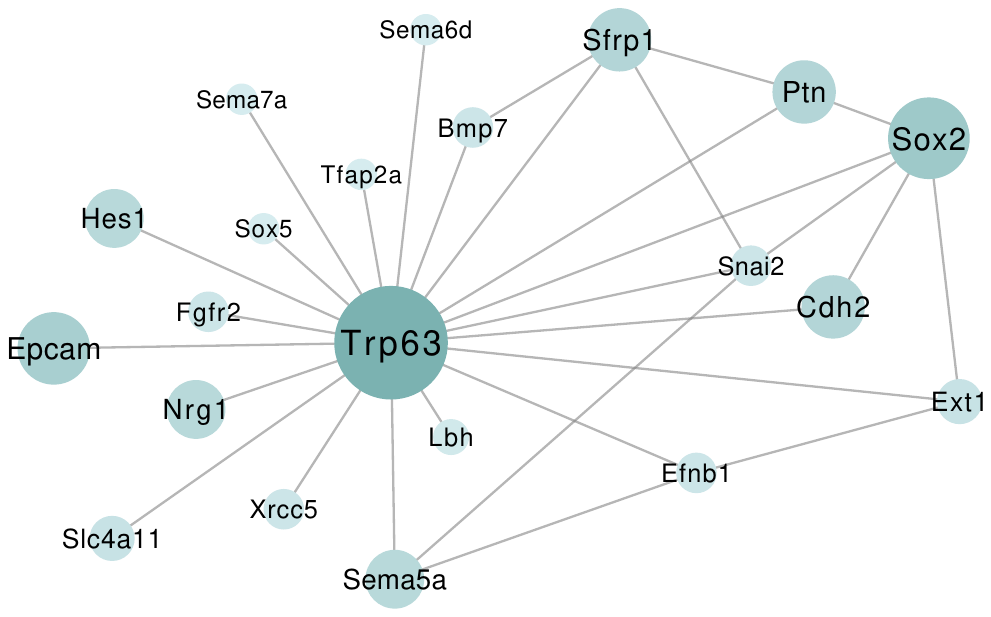}
    \vspace{-3cm}
    \caption{Sub-network of Trp63 extracted from Figure \ref{fig:HBCactstem}.}\label{fig:HBCactTrp63trans}
\end{figure}%

We want to stress that PC-zinb is able to find these putative TF-target pairs only on the basis of gene expression, hence proving itself as a useful tool to predict novel TF targets to be further validated with other techniques.

}


\section{Discussion}
\label{sec:discussion}

{ 
In this work, we have introduced PC-zinb, a class of constraint-based  algorithms for structure learning, supporting possibly overdispersed and zero-inflated count data. In focusing on these two nonstandard but realistic situations, our framework goes beyond what has so far been proposed in the literature. Moreover, by leveraging the proposal in \cite{JMLR:v22:18-401} -- shown to be competitive with state-of-the-art methods supporting count data -- we inherit the benefits of that approach, most notably: the existence of a theoretical proof of convergence of the algorithm under suitable assumptions;  an easy implementation of sparsity by a control on the number of variables in the conditional sets; invariance to feature scaling.
On the synthetic datasets considered in Section \ref{sec:sim}, we showed that the algorithms work well in terms of reconstructing the structure from given data for large enough sample sizes, while providing biologically coherent information and insight on the real dataset analyzed in Section \ref{sec:realdata}.

Our simulation studies  allow us to derive various recommendations on the use of PC-zinb. Clearly, these  do not  rule out sensitivity analyses with respect to both model specification and  tuning of the algorithms, which remain an important part of the model criticism process.
A control of the level of significance of the tests with respect to the sample size, $n,$ and the expected size of the neighborhood of each node, $b,$ is highly recommended to guarantee good reconstruction abilities. As in real applications knowledge of the expected size of the neighborhood might be difficult to elicit, it may be prudent to try a range of values for $b$, and check stability of results. This might also generate a sequence of models of  decreasing complexity for increasing values of $b$ and whose dynamic might also point researchers to the most significant connections.

If only the structure of the graph is of interest, irrespective of the strength of the links among variables, we suggest making use of the Poisson variant of the algorithm when the mean of the variables is small, so as to reduce computational complexity (Supplementary Tables S15 and S16). Moreover, when the mean of the variables is small, presence of zero-inflation might not influence reconstruction abilities of the algorithms, as also confirmed by the small study on zero-inflation in Section 5. In these situations, we recommend using, at least in the first instance, non zero-inflated models.

{
Clearly, in many applications, learning the structure might not be the only goal, and one might want to gain a quantitative insight into the dependence structure of the underlying process, by measuring the sign and the strength of the relations pictured in the graph. If the distribution needs also to be explicitly estimated, this can be achieved by using any of several existing parameter estimation methods conditional on the fixed structure learned by our approach.
}

If the null hypothesis $H_0: \beta^{\mu}_{st|K}=\beta^{\pi}_{st|K}= 0$ fails to be rejected, PC-zinb will remove the edge between variables $s$ and $t$. While such a procedure can only be justified in settings with high power, our simulation study shows that, even in settings with small sample sizes, our algorithm is able to achieve high power, and the correct underlying structure of the graph can be learned successfully.

While it is straightforward to interpret the case in which the neighborhood of $s$ is defined by the predictors of $\mu_{s|K}$, i.e., gene dependencies act on the average gene expression, the case of structure on $\pi_{s|K}$ requires more thought. If zero inflation represents true biological signal, we can interpret a non-zero $\beta^{\pi}_{st|K}$ as the fact that the presence of gene $t$ will influence the presence of gene $s$, regardless of their average expression. This is similar to \cite{mcdavid2019graphical}.
If zero inflation represents only technical noise, a simpler model with constant $\pi_{s|K}$ might be preferable. This is a special case of our general model. Since it is unclear what is the true nature of zero inflation in scRNA-seq data, we opted for generality in our model specification. Furthermore, having a general model expands the set of applications in which our approach may be useful.

\textcolor{black}{The question of whether zero-inflated models are useful for the analysis of scRNA-seq data has been frequently posed in the recent literature. In Section \ref{sec:ozi} we try to shed some light on why a negative binomial distribution can fit UMI data well, as observed by \citet{svensson2020droplet} and \citet{sarkar2021separating} among others (see also our Figure 3). We show that in the case of low mean and high variance the zinb and NB distributions are very close to each other, rendering the question of whether UMI data are zero inflated not. However, we also show that in real data zinb and NB models lead to different results, albeit with decent concordance between the inferred graphs (Supplementary Table S14). This result is only partially in agreement with those of \citet{sarkar2021separating}, in which the authors found that only a small percentage of genes show evidence of zero inflation. However, while \citet{sarkar2021separating} focus much of their attention to the case of univariate gene expression, modeling zero inflation may be important when looking at correlation between genes \citep[see also][]{yang2021modeling}.
}

{\color{black} Latent or unmeasured variables might induce associations between observed variables that can be spurious. Theoretical proposals are available to deal with the issue of latent factors in the setting in which the latent and observed variables are jointly Gaussian with the conditional statistics of the observed variables conditioned on the latent variables being specified by a graphical model \citep{Chandrasekaran}, but, to the best of our knowledge, no similar results are available for other families of models. For this reason, in our paper, we simply leverage on convergence of the PC algorithm to the model marginalized over the latent factors.}

{\color{black}As for the treatment of observed covariates and/or confounding factors, our proposed PC algorithm – that decomposes the structure learning problem into a series of tests performed on conditional log-likelihoods of each variable conditional on  other variables – naturally allows to incorporate the covariates into the conditional regression models and, therefore, to  estimate a covariate-adjusted structure for the graph. However, challenges remain if introduction of covariates augment the dimension of the conditional regression models to the point that one needs to resort to penalized tools. The treatment of both observed and latent covariates will be the object of future work.
}

Our real data analysis, aimed at assessing biological validity of the reconstructed network, has demonstrated the great importance of finding meaningful visualizations of large complex networks. Our proposal, based on a search for communities of variables and their association to gene ontologies via enrichment analysis, allowed us to confirm both biological interpretability of the estimated structure, and to contribute to our understanding of which and where biological processes are occurring.

}

\section{Software}
\label{sec5}

The methods presented in this article are available in the \textit{learn2count} R package, available at \url{https://github.com/drisso/learn2count} and as a zip file in the Supplementary Material \citep{supp}. The code to reproduce the analyses of this paper is available at \url{https://github.com/drisso/structure_learning} and as a zip file in the Supplementary Material \citep{supp}.






%
%
\section*{Acknowledgments}

The authors would like to thank Diya Das, Rebecca Chance, and John Ngai for providing access to the data and for help with the biological interpretation of the results.

DR was supported by ``Programma per Giovani Ricercatori Rita Levi Montalcini'' granted by the Italian Ministry of Education and University Research and by the National Cancer Institute of the National Institutes of Health (U24CA180996). TKHN was supported by the project of excellence ``Statistical methods and models for complex data'' awarded to the Department of Statistical Sciences, University of Padova by the Italian Ministry for Education and University Research. KVDB was a postdoctoral fellow of the Belgian American Educational Foundation (BAEF) and was supported by the Research Foundation Flanders (FWO), grant 1246220N. This work was supported in part by CZF2019-002443 (DR and TKHN) from the Chan Zuckerberg Initiative DAF, an advised fund of Silicon Valley Community Foundation.

\begin{supplement}\label{sec6}
\stitle{Supplementary material}
\sdescription{Supplementary material includes proofs and mathematical details, details on the real data analysis, and supplementary figures and tables. }
\end{supplement}

\begin{supplement}
\stitle{learn2count package}
\sdescription{R package with the implementation of the methods proposed in this article.}
\end{supplement}

\begin{supplement}
\stitle{Analysis code}
\sdescription{Code to reproduce the analyses of this article.}
\end{supplement}

\bibliographystyle{imsart-nameyear} 
\bibliography{refs}


\end{document}